\documentclass[conference]{IEEEtran}
\IEEEoverridecommandlockouts
\usepackage{cite}
\usepackage{amsmath,amssymb,amsfonts}
\usepackage{algorithmic}
\usepackage[ruled,vlined]{algorithm2e}
\usepackage{graphicx}
\usepackage{subfigure}
\usepackage{textcomp}
\usepackage{xcolor}
\usepackage{verbatim}
\usepackage{booktabs}
\usepackage{array}
\usepackage{caption}

\def\BibTeX{{\rm B\kern-.05em{\sc i\kern-.025em b}\kern-.08em
    T\kern-.1667em\lower.7ex\hbox{E}\kern-.125emX}}

\begin{document}

\title{  Large Language Models for Networking: Applications, Enabling
 Techniques, and Challenges  \\
}

\author{
	\IEEEauthorblockN{~Yudong~Huang,~Hongyang~Du,~Xinyuan~Zhang,~Dusit Niyato,~\IEEEmembership{Fellow,~IEEE},\\Jiawen~Kang,~Zehui~Xiong,~Shuo~Wang,~and~Tao~Huang,~\IEEEmembership{Senior~Member,~IEEE}\\  }

	\thanks{ Y. Huang and X. Zhang are with the State Key Laboratory of Networking and Switching Technology, BUPT, Beijing, 100876, P.R. China  (e-mail:
		hyduni@bupt.edu.cn, zhangxinyuan0181@bupt.edu.cn).
		
		H. Du and D. Niyato are with the School of Computer Science
		and Engineering, Nanyang Technological University, Singapore (e-mail:
		hongyang001@e.ntu.edu.sg, dniyato@ntu.edu.sg).
		 
		 J. Kang is with the School of Automation, Guangdong University of Technology, China (e-mail: kavinkang@gdut.edu.cn).
		 
    	Z. Xiong is with Information Systems Technology and Design (ISTD) Pillar, Singapore University of Technology and Design, Singapore (email: zehui\_xiong@sutd.edu.sg).
		
		 S. Wang and T. Huang are with the State Key Laboratory of Networking and Switching Technology, BUPT, Beijing, 100876, P.R. China, and the Purple Mountain Laboratories, Nanjing, 211111, P.R. China (e-mail: shuowang@bupt.edu.cn, htao@bupt.edu.cn).
		
}}

\maketitle

\begin{abstract}
 The rapid evolution of network technologies and the growing complexity of network tasks necessitate a paradigm shift in how networks are designed, configured, and managed. With a wealth of knowledge and expertise, large language models (LLMs) are one of the most promising candidates. This paper aims to pave the way for constructing domain-adapted LLMs for networking.  Firstly, we present potential LLM applications for vertical network fields and showcase the mapping from natural language to network language. Then, several enabling technologies are investigated,  including parameter-efficient finetuning and prompt engineering. The insight is that language understanding and tool usage are both required for network LLMs. Driven by the idea of embodied intelligence, we propose the ChatNet, a domain-adapted network LLM framework with access to various external network tools. ChatNet can reduce the time required for burdensome network planning tasks significantly, leading to a substantial improvement in efficiency. Finally, key challenges and future research directions are highlighted.
\end{abstract}

\begin{IEEEkeywords}
Large Language Models,  Generative AI, Intent-driven Networking, Network Intelligence.
\end{IEEEkeywords}

\section{Introduction}

Generative artificial intelligence (AI) technology is regarded as one of the most  inspiring  breakthroughs in the intelligent era. Through outstanding reasoning, generalization, and emergent abilities, large language models (LLMs) with billions of model parameters have shown great commercial value and technical potential, such as text-to-text, text-to-image, and text-to-code. The ChatGPT gains 1 million users within just one week, and open source LLMs (e.g., GPT-2, LLaMA, and BLOOM) are emerging one after another.

In particular, domain-adapted LLMs have been successfully utilized in robot embodied intelligence\footnote{Google PaLM-E: https://palm-e.github.io/}\footnote{ROS-LLM: https://github.com/Auromix/ROS-LLM}, chip design\footnote{NVIDIA ChipNeMo: https://research.nvidia.com/publication/2023-10\_chipnemo-domain-adapted-llms-chip-design}, and protein structure generation\footnote{AlphaFold: https://github.com/google-deepmind/alphafold}\footnote{ESM-2: https://github.com/facebookresearch/esm}. Generative LLMs can compress information features and vectorize massive knowledge as tokens, thereby aiding or even replacing humans in conceptual understanding, logical reasoning, and decision-making. Intuitively, this makes it possible to efficiently complete network tasks through natural language interaction with intelligent machines, while implementing domain-adapted  LLMs for vertical networking fields becomes an important research challenge.

Before the birth of LLMs, many research efforts trained task-specific AI models to express the paradigm of intent-driven networking. For instance, by leveraging a sequence-to-sequence learning model, a chatbot named Lumi\cite{lumi} was proposed to extract entities from the operator utterances, where  these entities are further translated into network intent language and deployable network policies. To reduce the configuration complexity of Access Control List (ACL) rules,  Language for ACL Intents (LAI)\cite{lai} was designed with specific grammar that contains three parts of region, requirement, and command.  In \cite{nassim}, the authors realized automated management of heterogeneous vendor-specific devices. It adopts the Bidirectional Encoder Representations from Transformers (BERT) model and learns directly from various devices' manuals to produce unified network data models.

Although these schemes perform well in certain network tasks and scenarios, there are several limitations: \textbf{1) Lack of generalization.} An AI model trained on a specific dataset may perform poorly on new or unseen network tasks. The lack of generalization ability prevents the AI model from being deployed in real network scenarios. \textbf{2) Huge training costs.} Training takes days or even months, as well as huge computing resources and labor costs, making it uneconomical to build an AI model from scratch. \textbf{3) Hard to integrate.} Existing intent-driven methods are limited to semantic conversion, which struggles to integrate with a wide range of off-the-shelf techniques (e.g., network simulator and search engine) and tools (e.g., solver, code interpreter, and visualization platform).

\begin{figure*}[]
	\centering
	\includegraphics[width=7in]{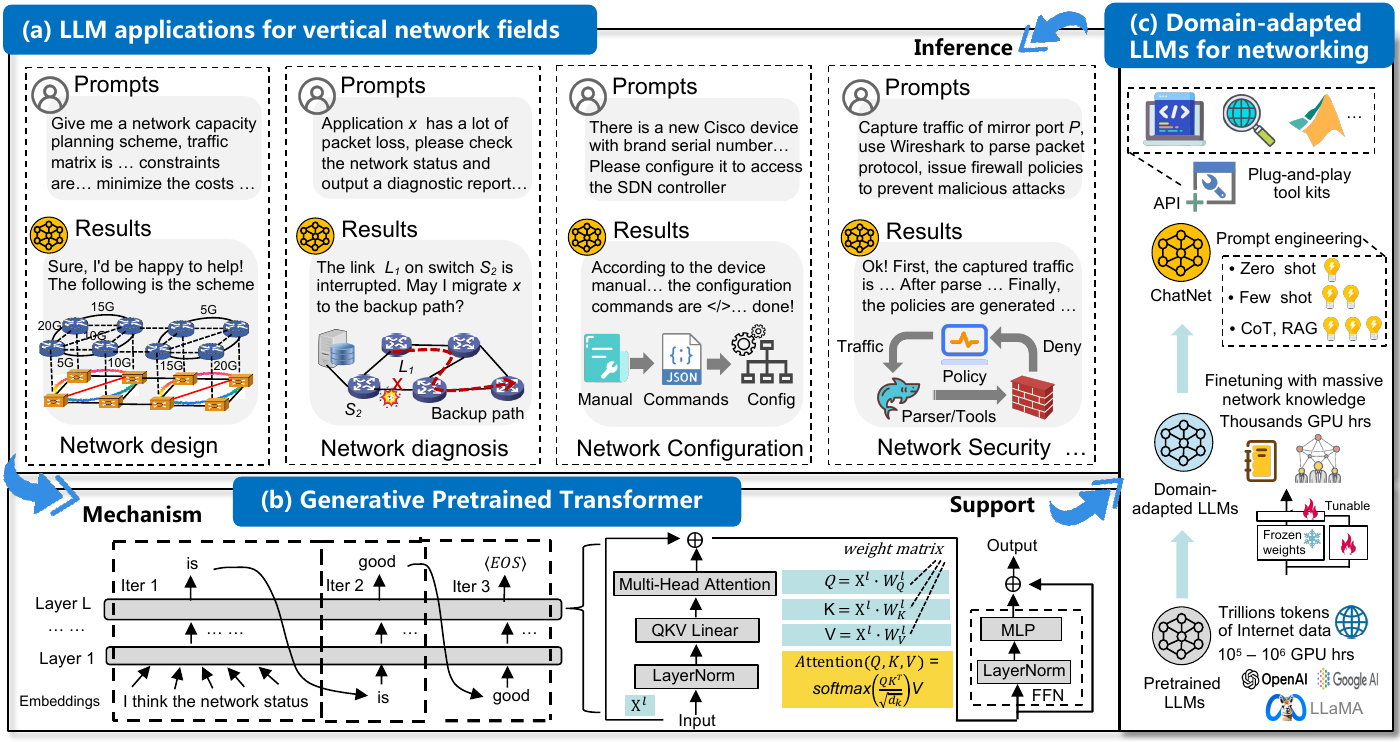}
	\caption{Applications, mechanisms, and enabling techniques for domain-adapted network LLMs. (a) Potential network LLM applications. (b) Working mechanisms of Generative Pretrained Transformer. (c) Finetuning process and prompt engineering. }
	\label{fig:LLM_NET}
\end{figure*}

The key point is that network language exists formal rules, protocols, mathematical expressions, and formula constraints, rather than plain text of natural language. Fortunately, LLMs are expected to continuously learn updated world knowledge and comprehensively utilize tools through application programming interface (API). Thus, arbitrary complex network tasks could be completed by calling LLMs with a combination of plug-and-play functional components. In this paper, we target to pave the way for constructing domain-adapted LLMs for networking, including applications in network design, network diagnosis, network configuration, and network security. We envision that LLM-based network intelligence will be ubiquitous and reshape future network infrastructure. The main contributions of this article are:

 $\bullet$ We analyze the features of natural language and network language, and showcase typical intent conversion patterns.

 $\bullet$ We present the enabling techniques of domain-adapted LLMs for networking, including pre-training, finetuning, inference, and prompt engineering.
 
  $\bullet$ We propose a conceptual framework, named ChatNet, with essential components  of analyzer, planner, calculator, and executor, to express the LLM-based network intelligence.
  
 $\bullet$ We conduct a case study of LLM-based network planning, where ChatNet can understand intents and generate visual capacity schemes with changing traffic matrices and constraints.

The rest of the article is organized as follows. We commence with the applications of LLMs for networking,  and exhibit the instances of network language. Next, we analyze the enabling techniques of domain-adapted LLMs for networking. Then the key functional components of ChatNet are detailed. Following that, we give the case study and analyze the challenges. Finally, we draw the main conclusions.

\section{Applications of LLM in Networking}

This section overviews the promising LLM applications for vertical network fields, emphasizing the discrepancy between natural language and network language.

\subsection{Potential LLM Applications for Vertical Network Fields}

Developing network systems and managing network infrastructure are knowledge-intensive and labor-intensive industries, which necessitate a lot of expert experience and manual operations. Previously, network intelligence was fragmented, residing in disparate small models, such as Deep Neural Network, Long Short-Term Memory, and Deep Reinforcement Learning. Each model was independently deployed within specific environments, such as intelligent assistants for customer service, adaptive routing algorithms for improving quality of service (QoS), and definite configuration synthesis modules for alleviating manual errors. LLMs promise to unify network intelligence through common natural language interfaces, making the network itself a generalist to understand the knowledge and master the tools. As shown in Fig. \ref{fig:LLM_NET}(a), we classify potential applications of LLMs in network vertical fields as follows.

\begin{table*}[h!]
	\centering
	\captionsetup{justification=centering}
	\caption{Mapping of natural language intents to typical network language implementations}
	\label{tab:MNL}
	\begin{tabular}{>{\raggedright\arraybackslash}p{5cm} >{\raggedright\arraybackslash}p{3cm} >{\raggedright\arraybackslash}p{3cm} >{\raggedright\arraybackslash}p{5.5cm}}
		\toprule
		\textbf{Natural Language (Intent)} & \textbf{Network Language} & \textbf{Characteristics} & \textbf{Examples of Mapping Relationships} \\
		\midrule
		``Restrict access to the server at 192.168.1.5 from all external IPs'' & Access Control List (ACL) & Control traffic based on IPs, protocols, ports, etc. &  \texttt{deny ip any 192.168.1.5}. \\
		\addlinespace[0.1cm]
		``Set up the new router to prioritize VoIP traffic for better call quality'' & Command Line Interface (CLI) / Policy & Configure network devices &  \texttt{class-map VOIP},\qquad\qquad \qquad\texttt{policy-map VOIP-Policy} \\
		\addlinespace[0.1cm]
		``Automatically adapt to changes in topology without manual reconfiguration'' & YANG Model / XML / JSON& Define data structure for network management. &   \texttt{$\langle$interface$\rangle$ \qquad\qquad\qquad\qquad$\langle$name$\rangle$10GE 1/0/1$\langle$/name$\rangle$} \\
		\addlinespace[0.1cm]
		``Parse all TCP packets and detect malicious and spoofed connections''  & Protocols (e.g., TCP, UDP, IP) & Define secure data exchange rules &  \texttt{IP header| TCP header| Payload} \\
		\addlinespace[0.1cm]
		``Ensure the network does not exceed 80\% capacity during 9 AM to 5 PM'' & Mathematical Formulas and Constraints & Constraints manage network performance &\texttt{if (time$\geq$ 9 AM and time $\leq$ 5 PM) { max\_load $\leq$ 0.8 * total\_capacity }} \\
		\bottomrule
	\end{tabular}
\end{table*}

\textbf{1) Network Design:} By processing vast datasets encompassing network performance metrics, equipment specifications, and historical design patterns, LLMs can assist engineers in equipment selection, network planning, protocol  formulation, and many other aspects of network design\footnote{LossLeaP for traffic prediction and capacity forecasting: https://github.com/alcoimdea/LossLeaP}. In equipment selection, LLMs could analyze compatibility requirements, performance benchmarks, and cost considerations, providing recommendations that align with specific network objectives. For network planning, LLMs may simulate various network schemes, predict potential bottlenecks, and suggest optimal layouts that balance efficiency, scalability, and resilience.

\textbf{2) Network Diagnosis:} Troubleshooting is a tedious and burdensome task for network operators. Especially in large-scale wide-area networks, it requires coordination between different departments across multiple regions, while applications still suffer from inexplicable network failures or performance degradation, and are threatened with hundreds of millions of financial losses. By integrating LLMs into network diagnostic systems\footnote{Juniper Marvis: https://www.juniper.net/us/en/products/cloud-services/virtual-network-assistant.html}, LLMs are capable of generating fault reports based on network status information, accelerating fault location, and giving reasonable processing suggestions based on the report analysis and historical operational data.

\textbf{3) Network Configuration:} There are a large number of heterogeneous devices in the network, e.g., switches, routers, and middleware. Due to vendor-specific device models, significant expert effort is required to learn the user manuals, collect suitable commands, validate configuration templates, and map template parameters to the controller database. In this process, even a single ACL misconfiguration may lead to network disruptions. Considering the growing heterogeneous cloud networks with plenty of computing and storage devices that also need to be managed, a unified natural language configuration interface\footnote{Huawei NAssim: https://github.com/AmyWorkspace/nassim} is essential for simplifying the configuration process and enabling self-configured networks.

\textbf{4) Network Security:} Networks often face various potential security issues, such as distributed denial-of-service (DDoS) attacks, address spoofing, and data leakage. Protecting the network from malicious attacks  combines a series of operations, such as security assessment, vulnerability scanning, intrusion detection and defense. LLMs are powerful interactive platforms to access diverse security tools and systems\footnote{Google Cloud Security AI Workbench and Sec-PaLM 2: https://cloud.google.com/security/ai}. For instance, guided by logically rigorous prompts, LLMs may complete the abnormal traffic denying tasks by calling the parse tool of Wireshark and updating the policy to firewalls.

\subsection{From Natural Language to Network Language}

As depicted in Table \ref{tab:MNL}, different from the plain text of natural language, the network language contains more non-standardized formats and symbols, from the high-level management policy to the low-level Access Control List (ACL), Command Line Interface (CLI), and data modeling language (e.g., YANG model, XML, and JSON). Traditional mapping methods are restricted to formalized translations, e.g., entity abstraction and template filling. In contrast, LLMs can provide better network QoE performance by offering customized responses to specific human-related natural language inputs. Moreover, the network language has domain-specific nouns, protocols, and rules, as well as mathematical constraints, where LLMs are prone to ``illusion" due to ambiguous concepts or ``babbling” due to forgetting the relevance of the context. \textit{Thus, our first insight is that we can finetune the LLMs with massive network knowledge to enable domain-adapted network LLMs, and Retrieval Augment based on accessing external documents (e.g., device manuals and status logs) will benefit mapping the natural language to the network language.}

\begin{table*}[]
	\centering
	\captionsetup{justification=centering}
	\caption{An overview of studies for LLMs with intent-driven networking (IDN) and network embodied intelligence (NEI).}
	\label{tab:compare}
	\begin{tabular}{>{\raggedright\arraybackslash}p{1.8cm} >{\raggedright\arraybackslash}p{8cm} >{\raggedright\arraybackslash}p{1cm} >{\raggedright\arraybackslash}p{1cm} >{\raggedright\arraybackslash}p{1.2cm} >{\raggedright\arraybackslash}p{1cm}>{\raggedright\arraybackslash}p{1cm}}
		\toprule
		\textbf{Article} & \textbf{Contributions} & \textbf{IDN} & \textbf{LLMs} &\textbf{Finetuning}&\textbf{Prompt}&\textbf{NEI}\\
		\midrule
		LLM for wireless\cite{bariah} & Introduce applications of LLMs in future wireless communications, including designing,
		training, testing, and deploying Telecom LLMs.&\ &  $\surd$&  $\surd$ & \ & \ \\
		\addlinespace[0.1cm]
		AIGN\cite{aign} & A generative network system that can generate  customized network solutions with the diffusion model-based learning approach.&$\surd$&$\surd$&  \ & \ & \ \\
		\addlinespace[0.1cm]
		NetGPT\cite{netgpt} &   A collaborative cloud-edge methodology towards personalized LLM services and native-AI network architecture.& \ &$\surd$&  $\surd$ & $\surd$  & \ \\
		\addlinespace[0.1cm]
		Ours ChatNet & Introduce domain-adapted LLMs for vertical network fields and propose the network embodied intelligence. & $\surd$& $\surd$&   $\surd$& $\surd$& $\surd$\\
		\bottomrule
	\end{tabular}
\end{table*}

In addition, completing network tasks are complex and error-prone process, which requires not only semantics correctness but also being practically deployable. To address this issue, the intent refinement\cite{refinement} was proposed to guarantee the
accuracy and completeness of the translation from the declarative intent to network primitives, with methods like Bi-LSTM and knowledge-graph. The network verification\cite{verification} further checks the conflicting policies and validates the feasibility of results with various network tools. However, assembling disparate tools and approaches in intent-driven networking is still challenging.  In this paper, we consider the generative pre-trained transformer (GPT) techniques of LLMs to be powerful enough to learn the usage of tools, as GPT-4 has released the assistant APIs\footnote{OpenAI assistants API: https://platform.openai.com/docs/assistants/overview} for easily constructing customized GPT applications. \textit{Thus, our second insight is that domain-adapted network LLMs can access external tools\cite{gao2023assistgpt}, such as search engines, data analyzers, mathematical solvers, and network tools, to automate any complex tasks, such as processing time series data, parsing protocols, and constructing mathematical models. Then, we call these kinds of network LLMs as ChatNet.}

\section{Domain-adapted  LLMs for Networking}
Motivated by the above issues and trends, this section analyzes the enabling techniques for network LLM, especially specific design problems for LLMs to make the LLM models optimized and suitable for network applications, and several assessment methods in prompt engineering are derived.

\subsection{Enabling Techniques for Network LLMs}

We divide the enabling techniques for network LLMs into three categories: pre-training, finetuning, and inference. The pre-training part presents the working mechanism of LLMs. Based on open-source vanilla LLMs, finetuning is the most important step in establishing domain-adapted network LLMs.

\textbf{1) Generative Pre-trained Transformer:} The training of LLMs involves large-scale unsupervised learning, where the model is pre-trained on extensive text corpora ($\sim$PBs) for $10^{5}\sim10^{6}$ GPU hours, learning to predict the next word in a sentence. As shown in Fig. \ref{fig:LLM_NET}(b),  the input is word embeddings, which goes through all layers of the GPT. Each layer mainly consists of a multi-head attention module and a position-wise feed-forward network (FFN). The multi-head attention mechanism utilizes multiple attention processes to simultaneously computes the relevance of each word in a sentence to every other word, using a set of Query ($Q$), Key ($K$), and Value ($V$) vectors and the intermediate result $x^{l}$.  The outputs of these attention heads are then concatenated and linearly transformed to produce next-token probabilities. A single training run would cost millions of dollars, while each thousand API calls cost less than one dollar for users.

\textbf{2) Parameter-efficient Finetuning:} Since training LLMs from scratch is prohibitively costly and calling third-party LLMs via APIs poses data security risks, finetuning open-source LLMs is a viable candidate for building domain-adapted network LLMs. As shown in Fig. \ref{fig:LLM_NET}(c), parameter-efficient  finetuning refers to freezing most of the pretrained weights to adjust specific layers or adding additional tunable parameters. For instance, Low-Rank Adaptation (LoRA) introduces low-rank matrices to approximate the changes needed in the model's weights. Compared to full-parameter approaches, LoRA strikes a balance between maintaining the general capabilities of the original model and adapting it to specific network domains and network tasks. 

Particularly, the success of finetuning largely depends on the quality of data source and instruction datasets. Network data collection encompassed a variety of operational reports, user manuals, scripting languages, protocol descriptions, debugging logs, and configuration files. Standardizing the datasets includes a series of processes, such as cleaning, filtering, categorization, normalization, and anonymization. Moreover, a domain-adapted tokenizer can be trained to improve the tokenization efficiency, by adding new tokens for domain-specific network terms, such as keywords commonly found in network protocols. To perform supervised finetuning on the domain-adapted network LLMs, high-quality network instruction datasets need to be  established for specific network tasks.

\textbf{3) Context-aware Inference:} More than a mapper from natural language to network language, domain-adapted network LLMs have unprecedented in-context learning and multi-turn dialogue capabilities. Context-aware inference means that we can continuously provide new prompts to guide LLMs to perform logical reasoning, avoiding the repeated construction of preconditions and environments. Recently, the GPT-4 Turbo model has supported a context window of 128K tokens, where more than 300 pages of text can be fitted in a single prompt. This opens up possibilities for LLM-based network applications with massive information analysis and processing. The ``Emergent'' abilities of LLMs  have great potential to intelligently create novel designs, mechanisms, and protocols even unseen in existing network environments.  

\begin{figure*}[]
	\centering
	\includegraphics[width=7in]{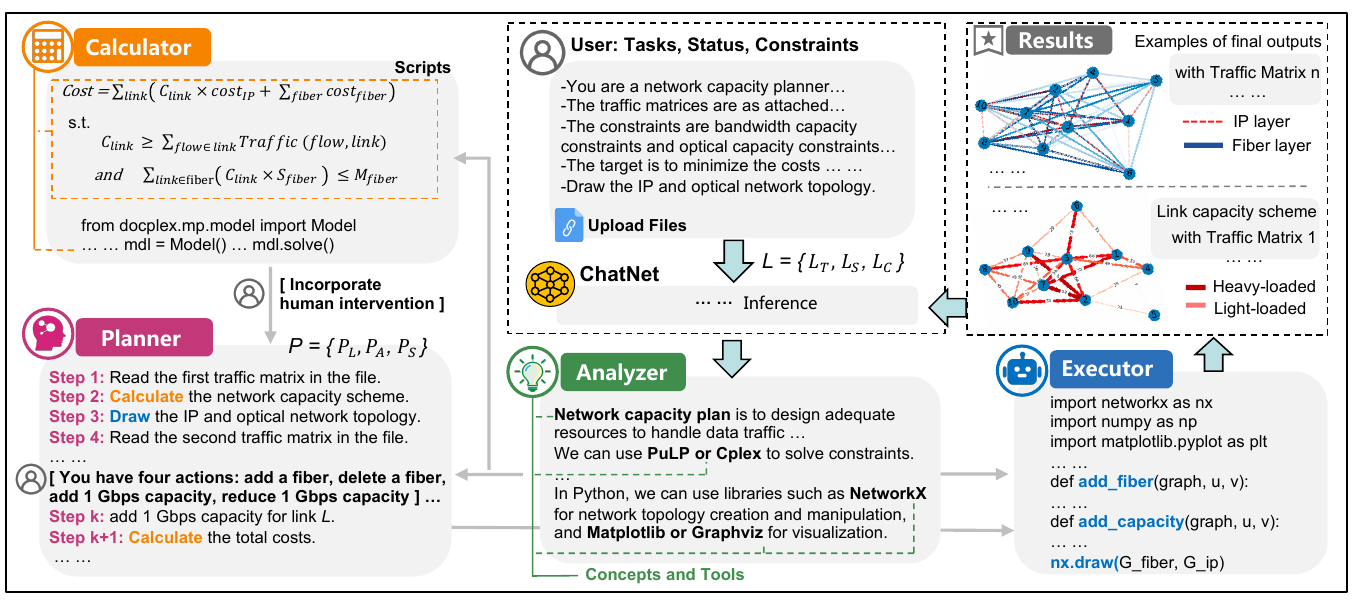}
	\caption{The ChatNet consists of the analyzer, planner, calculator, and executor, each of which is powered by a network LLM. Under the case study of network planning, ChatNet is fed with prompts and ultimately outputs diverse capacity schemes.}
	\label{fig:ChatNet}
\end{figure*}

\subsection{Prompt Engineering for Assessment}

Effective prompt engineering is crucial for maximizing the potential of network LLMs in various applications, involving crafting questions or statements that guide the LLMs to produce the desired output. Especially for complex network tasks, the prompt should be clear with sufficient context, such as satisfying specific constraints and response formats. Multiple choice questions are a widely accepted assessment format for LLMs. The authors of \cite{tele} collected a TeleQnA dataset with 10000 questions and answers, serving as an evaluation tool for assessing the knowledge of LLMs in the telecommunications domain. In \cite{emp}, a NetEval dataset with 5732 questions was constructed to measure the comprehensive capabilities of 26 publicly available LLMs in network operations. The following are several commonly used prompt methods.

\textbf{1) Zero Shot and Few Shot:} Zero-shot prompts only contain the task description and test questions, which are suitable for simple networking tasks, such as explaining concepts in technical specifications. The few-shot prompts are slightly more complex. It includes a small set of examples (usually one to three) that demonstrate the desired task or answer format. These examples serve as a mini-training set, guiding the model to understand the context and the specific nature of the response required. In the network fields, these examples might be brief descriptions of network configurations, troubleshooting scenarios, or protocol interactions, followed by the query that requires a similar response.

\textbf{2) Chain of Thought:} Chain-of-Thought (CoT)\cite{chain_of_t} enables LLMs to tackle complex arithmetic, commonsense, and symbolic reasoning tasks. This approach encourages the LLMs to follow a step-by-step reasoning process, akin to how a human might break down a problem. By using CoT prompting,  LLMs can be guided to systematically analyze a networking problem. For example, in diagnosing a network issue, the prompt can be engineered to lead the model through a series of diagnostic steps, considering various factors like network topology, hardware status, and software configurations. Moreover, CoT prompts can be designed to facilitate the cascading use of plug-and-play network tools,  empowering LLMs from mere answer generators into network experts who are capable of tools using, logical reasoning, and problem-solving.

\textbf{3) Retrieval-Augmented Generation:} Retrieval-Augmented Generation (RAG) combines LLMs with information retrieval techniques, notably vector retrieval, to enhance the model's memory and factual accuracy. When a query is received, the model first performs a semantic search across a vast database, selecting text pieces that are semantically relevant to the query. This selection is facilitated by semantic indexing, which efficiently organizes and retrieves data based on its meaning rather than just keyword matching. The retrieved information is then fed into the LLM as a part of the prompt, effectively providing the model with a context-rich background to generate more informed and accurate responses. With the RAG,  we can introduce up-to-date datasets (e.g., iterative standard drafts and updated maintenance logs) to LLMs dynamically, eliminating the need for constant retraining of LLMs with new samples.

\section{ChatNet Framework} 

After domain-adapted enhancement, it may still be a step away from real implementation of the ChatNet, as knowledge understanding alone does not directly derive the ability to use tools. This section provides an in-depth analysis of the key components required to improve proficiency in utilizing network tools and proposes the ChatNet Framework. Additionally, we compare ChatNet with other recent studies in Table \ref{tab:compare}, followed by a case study in network planning scenarios.

\subsection{ Essential Components of ChatNet}
The utilization of tools is a key indicator of advanced intelligence, as demonstrated in the behaviors of the human and the robot embodied intelligence\cite{creative}. In the same vein, the ChatNet should master both the language understanding and the tool usage, based on the following four fundamental modules of analyzer, planner, calculator, and executor.

\textbf{1) Analyzer:} Powered by network LLMs,  the analyzer is designed to extract key concepts, tools, and their relationships to  assess the feasibility of network tasks. Generally, the analyzer is fed with a prompt of natural language descriptions $L$ that should cover the range of $\left\{ L_{T}, L_{S}, L_{C} \right\} $, where $L_{T}$ is the task description, $L_{S}$ is the network state, and $L_{C}$ denotes network constraints. Moreover, there is an additional file interface for uploading datasets and linking files.

\textbf{2) Planner:} The planner reasons out the necessary step-by-step process to complete the network tasks, where the planning space $P$ is defined by sets $\left\{ P_{L}, P_{A}, P_{S} \right\} $. $P_{L}$ depicts the planning logic, such as simple sequence or loop steps. $P_{A}$ is a collection of customizable operations and actions, such as reading files and accessing tools. $P_{S}$ denotes the skills that are needed to leverage specific network tools. It is worth noting that the proposed module is serving as plan creation, rather than plan execution. This means that users can conduct multiple rounds of dialogue with the planner through the CoT, and even modify the plan directly.

\textbf{3) Calculator:}  LLMs are not good at network mathematic and formulation, while there are complex numerical calculations and model constraints in the network system. Thus,  LLMs must have an additional calculation module to compute parameters for each step. For instance, the calculator can invoke programming language to implement simple arithmetic operations, or import solvers to optimize constrained models. Network LLMs can generate useful scripts based on prompts to speed up the network modeling process. Some human intervention is inevitable for the collaboration of the calculator and the planner considering complex network tasks.

\textbf{4) Executor:} The executor is responsible for outputting the final results. Typically, the executor generates networking schemes and protocols, as well as network configuration commands (e.g., ACL and CLI) by coding. Through unified API of network LLMs, the executor can also be integrated into network emulators, controllers, and verification tools.

\subsection{Case Study under Network Planning}
We simulate a prototype of ChatNet with the support of GPT-4, where four GPT-4 models are initially prompted as analyzer, planner, calculator, and executor, respectively.  As illustrated in Fig. \ref{fig:ChatNet}, the prompts position the role of analyzer as a network planner, and then inform it of the traffic matrices, capacity constraints, optimization goals, and desired task outputs. Firstly, the analyzer explains network capacity planning and points out the required tools, such as Cplex for constraint solving, NetworkX for network topology creation, and Matplotlib for visualization. These outputs are manually delivered to other modules as input prompts. Then, the planner module decouples tasks and starts executing them step by step, which mainly includes reading traffic matrix files, calculating network capacity solutions, and drawing IP and optical network topology. Moreover, personalized actions (e.g., add the fiber or add the capacity) can be designated in the planner to instruct further modifications to the network topology. The calculation formula of the cost and optimization model are stored in the calculator module prior, which is a combination of scripts provided by ChatNet and human intervention. Finally, the executor generates customized network capacity solutions, such as using colors to display different congestion levels, and using dotted and solid lines to display hierarchical IP and optical network topologies.

\begin{figure}[]
	\centering
	\includegraphics[width=3.4in]{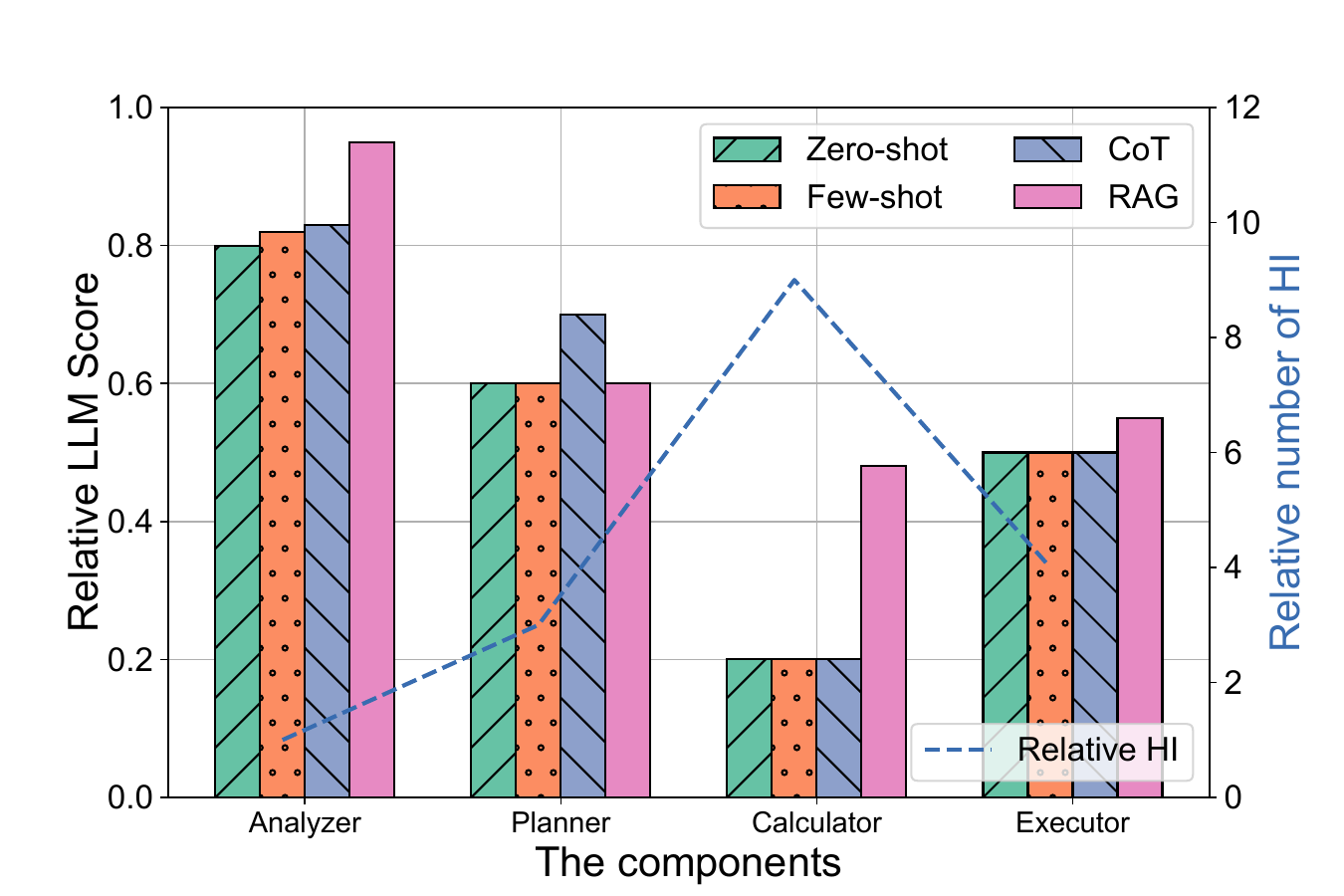}
	\caption{Four prompt methods of zero-shot, few shot, CoT, and RAG are compared.  The relative LLM score and number of human intervention (HI) are adopted to evaluate the performance of the four components of ChatNet.}
	\label{fig:sim_chatnet}
\end{figure}

Due to the lack of task benchmarks and instruction datasets in network domains, we combine the LLM evaluators\footnote{An instance of the LLM evaluator to score the output quality of the analyzer module is at https://github.com/Hyduni001/ChatNet.}\cite{wang2023chatgpt} and expert evaluation\cite{lee2023rlaif} to briefly assess the four components under the prompt methods of zero-shot, few shot, CoT, and RAG. The scoring is normalized to a range of $[0,1]$.  In this scale, scores of 0.8, 0.6, and 0.4 indicate that the results align with expectations to a high, moderate, and just adequate degree, respectively. A score of 0.2 signifies that the results do not match expectations. If automatic prompt delivery can be achieved between modules, more objective metrics could be used to evaluate the overall performance of ChatNet, such as high-level planning (HLP)\footnote{LLM-Planner: https://osu-nlp-group.github.io/LLM-Planner/} accuracy and the percentage of completed tasks.  As shown in Fig. \ref{fig:sim_chatnet},  we find that the analyzer always achieved a score higher than 0.8 with no more than one count of human intervention (HI), while the calculator is the bottleneck of the entire framework and requires multiple human interventions. When the RAG is employed by uploading a specific network planning design document, the calculator is capable of extracting mathematical constraints and providing a model that is basiclly correct. Moreover, the CoT can improve the performance of the planner to some extent.  It is worth noting that LLMs are prone to errors during the inference process, and currently do not match expert performance on network planning tasks. However, LLMs reduce the time required for burdensome network planning tasks, which greatly improves efficiency. Building extensive network tasks and instructions to evaluate a variety of domain-adapted network LLMs will be future work.

\section{Challenges and Future  Prospects}
In this section, we analyze the challenges brought by the domain-adapted network LLMs and highlight the potential research directions.

\subsection{Training Multi-modal Network LLMs}
The integration of diverse data types, such as text, images, and network-specific codes, requires a sophisticated training process to construct multi-modal LLMs for networking. The model must be adept at processing and interpreting this heterogeneous data in a way that accurately reflects the complexities of network environments. Moreover, there is an issue of maintaining model relevance over time. Network technologies and protocols evolve rapidly, necessitating continuous updates to the training data. Balancing these factors while minimizing training costs and computational resources is a significant challenge, which must be addressed to fully earth the potential of multi-modal network LLMs.

\subsection{Developing Network LLM Plugins}
The development of network LLM plugins opens a new frontier in network management and design. These plugins are intended to extend the capabilities of LLMs, allowing them to interact more effectively with various network components and systems. The challenge lies in designing plugins that are both flexible enough to accommodate a wide range of network architectures and specific enough to provide meaningful insights and actions. Interoperability is a key concern, as these plugins must seamlessly integrate with existing network management tools and protocols. Additionally, ensuring the security and reliability of these plugins is paramount.

\subsection{Enabling Network Embodied Intelligence}
 The realization of network embodied intelligence through LLMs holds the promise of more responsive, efficient, and self-optimizing network systems, representing a significant leap forward in network fields. For instance, it would be meaningful to integrate network LLMs into decision-making systems, such as the network planning system with the deep reinforcement learning. Furthermore, one fundamental issue is the transparency and explainability, since LLMs may create fake network designs and configurations. Similar to the intelligence from L0 to L5 in autonomous driving, network embodied intelligence needs to be considered in layers, e.g., from assisting the netowrk operators to completely replacing the network experts.

\section{Conclusions}

This article has studied the applications of LLMs for networking. We summarized the enabling techniques for establishing domain-adapted network LLMs and analyzed the prompt engineering of zero-shot, few shot, CoT, and RAG. A novel ChatNet framework is proposed to exhibit the network embodied intelligence and a case study has been conducted under the network capacity planning scenario. Finally, the challenges, such as training multi-modal network LLMs, and developing network LLM plugins, are discussed. We hope that ChatNet can serve as an inspiration for future research.

\bibliographystyle{IEEEtran}

\bibliography{IEEEabrv, reference.bib}

\end{document}